\documentstyle[twoside,fleqn,elsevier,epsf]{article}

\newcommand{\nwc}{\newcommand}

\nwc{\be}  {\begin{equation}}
\nwc{\ee}  {\end{equation}}
\nwc{\bmu} {\bar{\mu}}
\nwc{\ba}  {\begin{eqnarray}}
\nwc{\ea}  {\end{eqnarray}}
\nwc{\bi}  {\begin{itemize}}
\nwc{\ei}  {\end{itemize}}
\nwc{\nn}  {\nonumber\\}
\nwc{\Tr}  {\mathop{\rm Tr}}
\nwc{\im}  {\mathop{\rm Im}}
\nwc{\Hc}  {\mathop{\rm H.c.}}
\nwc{\eq}  {Eq.~}
\nwc{\fig} {Fig.~}
\nwc{\la}[1]{\label{#1}}
\nwc{\rmi}[1]{{\mbox{\scriptsize #1}}}
\nwc{\nr}[1]{(\ref{#1})}
\nwc{\fr}[2]{{\frac{#1}{#2}}}
\nwc{\msbar}{\overline{\mbox{\rm MS}}}

\def\lsi{\raise0.3ex\hbox{$<$\kern-0.75em\raise-1.1ex\hbox{$\sim$}}}
\def\gsi{\raise0.3ex\hbox{$>$\kern-0.75em\raise-1.1ex\hbox{$\sim$}}}
\nwc{\lsim}{\mathop{\lsi}}
\nwc{\gsim}{\mathop{\gsi}}

\title{What's new with the electroweak phase transition?%
       \thanks{Work partly supported by the TMR network 
       {\em Finite Temperature Phase Transitions in Particle
       Physics}, EU contract no.\ FMRX-CT97-0122.
       Presented  at {\em Lattice '98} 
       by M. Laine.}}

\author{M. Laine\address{Theory Division, CERN, CH-1211 Geneva 23,
        Switzerland} 
        and 
        K. Rummukainen\address{NORDITA, Blegdamsvej 17,
        DK-2100 Copenhagen \O, Denmark}} 

\begin{document}

\begin{abstract}
We review the status of non-perturbative lattice studies of the electroweak
phase transition. In the Standard Model, the complete phase diagram
has been reliably determined, and the conclusion is that there is no
phase transition at all for the experimentally allowed Higgs masses.  In
the Minimal Supersymmetric Standard Model (MSSM), in contrast, there can be 
a strong first order transition allowing for baryogenesis. Finally, we
point out possibilities for future simulations, such as the problem of
CP-violation at the MSSM electroweak phase boundary.
\end{abstract}
\maketitle

\vspace*{-9.3cm}

\begin{minipage}[t]{15.5cm}
\begin{flushright}
\mbox{hep-lat/9809045} \\
\mbox{September, 1998}
\end{flushright}
\end{minipage}

\vspace*{7.7cm}

\section{INTRODUCTION}

Primordial nucleosynthesis computations
tell that the net baryon to photon 
number ratio $\eta$ in the Early Universe is a definite
non-vanishing number, $\eta = (1...9) \times 10^{-10}$. 
On the other hand, it is natural to assume that after
inflation, $\eta=0$. The last instance in the 
post-inflationary history 
of the Universe during which $\eta>0$ could have been generated,
is the electroweak phase transition,
at $T_c\sim 100$ GeV~\cite{krs}.
Thus, electroweak baryogenesis is in a way the most conservative
scenario of baryon number generation, and at the same time, 
the only scenario which is experimentally testable
in existing collider experiments. 

The scenario of electroweak baryogenesis has quite 
a few different ingredients. To generate a baryon number, 
one needs anomalous baryon number violating processes
in the symmetric high temperature phase, microscopic
C- and CP-violation, and a thermal non-equilibrium (for 
a review, see~\cite{rs}). It is perhaps surprising
that many of these ingredients can be studied
non-perturbatively with lattice simulations. 
Indeed, baryon number violation has been studied 
both in the symmetric and broken phases of the theory
(\cite{moore_br,st} and references therein).
Something can perhaps also be said about CP-violation
in the MSSM (Sec.~\ref{sec:CP}). Finally, whether
there is non-equilibrium or not, depends on the order and
strength of the transition.

The main interest here will be on
the last of these questions. 
In addition to what kind of non-equilibrium phenomena there can be, 
this determines whether the baryon number violating processes
are sufficiently switched off after the transition for any 
baryon number possibly produced to remain there. Indeed, while
the baryon number violating rate is assumed to be only
parametrically suppressed before the transition, 
$\dot B\sim \alpha_W^5 \ln(C/\alpha_W) T^4$ \cite{db}, 
it is exponentially suppressed after the transition,
$\dot B\sim \exp[-45(v_H/T)]T^4$~\cite{krs}. 
The general constraint for the baryon number
generated to remain there after the
transition is $v_H/T_c \gsim 1$~\cite{krs,rs},
and we would hence like to compute this ratio.

\section{PERTURBATION THEORY}

In principle, $v_H/T$ can be computed in perturbation theory.
So why does one need simulations? This may seem like quite a
relevant question, as we are studying the electroweak
sector of the Standard Model, for which 
perturbation theory works perfectly at zero temperature.

However, things are different at finite temperatures. This is 
due to the so-called infrared problem. The point is simply that 
there is a new scale $T$, and there can thus be new
expansion parameters such as ${g^2T}/({4\pi m})$, where 
$g^2$ represents the couplings of the theory and $m$ the 
masses. In practice, this kind of an expansion parameter 
can emerge from the Bose-Einstein distribution $n_b$: 
\ba
g^2 n_b(m) = \frac{g^2}{e^{m/T}-1}
\stackrel{m\ll T}{\sim}\frac{g^2T}{m}.
\ea
Thus, it is the bosonic degrees of freedom which are
particularly problematic. Now, if $m$ represents, e.g., the
W mass $m_W = {gv_H}/{2}$, then $m\approx 0$ in the symmetric phase,
and perturbation theory need not work at all. The 
phase transition takes place between the symmetric and broken
phases, and thus its characteristics may also be 
unreliably described. Thus
we need lattice simulations.

\section{NON-PERTURBATIVE METHODS}

\subsection{4d simulations}

In principal, the most straightforward way to attack the problem
is to do standard four-dimensional
(4d) finite temperature lattice simulations. However,
in the present context
4d simulations turn out to be quite demanding. This is, somewhat
surprisingly, due to the fact that 
the coupling is weak. A weak coupling makes the system
have multiple scales, and if a lattice with spacing
$a$ and extent $N$ is to describe the infinite 
volume and continuum limits, one must require 
\be
a \ll \frac{1}{\pi T} \ll \frac{1}{\sqrt{2}g T} \ll 
 \frac{1}{g^2T} \ll  Na. \la{requirement}
\ee
For small $g$, lattices thus need to be very large.

It is quite remarkable that in spite of this severe requirement, 
a continuum extrapolation can sometimes be carried out (see below).
In particular, one can employ an asymmetric
lattice spacing~\cite{cfh1}, which should essentially
relax the leftmost inequality in \eq\nr{requirement}.

The second problem with the 4d simulations is that 
only the bosonic sector of the Standard Model
can be studied, since chiral quarks (especially the top)
cannot be put on the lattice. 

\subsection{3d simulations}

Another possibility is simulations in a three-dimensional 
(3d) effective theory. The main idea of this approach
is to combine the best parts of perturbation theory and simulations:
one can integrate out massive modes perturbatively, which works well
since the couplings are small, and then study light modes 
non-perturbatively. In the first step, the original 4d theory
reduces to a 3d one.

The 3d approach allows to overcome the two problems of
4d simulations mentioned above. Indeed, since the large mass scales 
($\pi T, \sqrt{2}gT$) are removed, it is easier to 
satisfy \eq\nr{requirement}.
Consequently, an infinite
volume and continuum extrapolation with a relative error
of, say, 5\%, can be obtained with rather moderate
computer resources (for a review, see~\cite{krreview}).
Moreover, one can study the 
full theory with chiral fermions~\cite{generic}, 
for realistic (small) gauge couplings. In fact, 
the same simulation results tell about the infrared
properties of many other theories, as well, such as the MSSM
in a part of its parameter space~\cite{mssm}. 

The main question concerning dimensional reduction is, 
of course, how accurate the effective theory constructed
really is. Parametrically, the error $\delta G$ for
static Green's functions $G$ is
${\delta G}/{G} \lsim O(g^3)$~\cite{generic}, but
an essential point is to convert this parametric estimate
into a numerical one. What one can do is to compute
some particular higher dimensional operators, and see
what kind of an effect they give in different infrared
Green's functions. This leads to errors 
on the $\sim 1\%$ level, and often even smaller in theories without 
the top quark~\cite{generic}. 
A conservative error estimate is then 
$\lsim 5$\% in the Standard Model (for $m_H\lsim 250$ GeV).
In the MSSM the errors can be a bit larger, since the strongly
interacting squarks may play a significant role,
and since there are more mass parameters which may compromise
the high temperature expansion. It is perhaps worth stressing that 
perturbative dimensional reduction does not work at all 
for the QCD phase transition, where the coupling is large.

\section{SIMULATION RESULTS}

We now review the main recent 
physics results obtained with the two approaches.

\subsection{The Standard Model}

In the 3d approach, the 
Lagrangian relevant for the Standard Model is
\be
{\cal L}_{\rm 3d} = 
{1\over4} F^a_{ij}F^a_{ij}+
(D_i\phi)^\dagger D_i\phi+m_3^2\phi^\dagger\phi+
\lambda_3(\phi^\dagger\phi)^2. \la{SMlagrangian}
\ee
The U(1) group has here been neglected (i.e., $\sin^2\theta_W=0$),
since its effects are small~\cite{su2u1}. Let us 
denote 
\be
x=\lambda_3/g_3^2,\quad y=m_3^2(g_3^2)/g_3^4.
\ee
In the 4d simulations, one studies the \linebreak
SU(2)+Higgs theory, whose Lagrangian is precisely
\eq\nr{SMlagrangian} but in 4d.

\begin{figure}[tb]

\vspace*{0.2cm}

\epsfxsize=7.3cm
\centerline{\epsffile{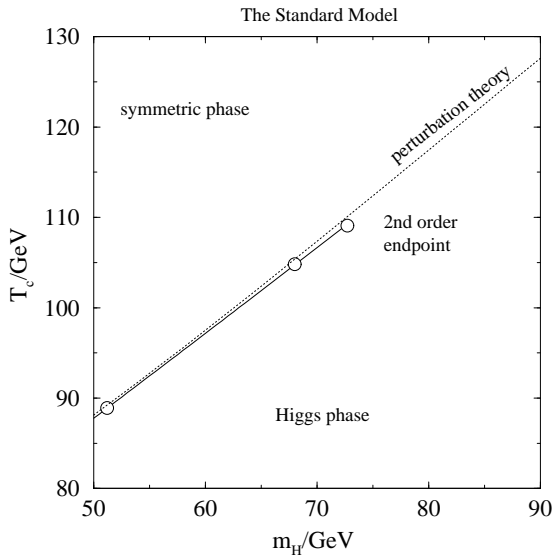}}
 
\vspace*{-1cm}

\caption[a]{The phase diagram of the Standard Model.
The non-perturbative endpoint location has been studied
with 3d simulations in~[11--14]
and with 4d simulations in~[15--18].
In perturbation theory (dotted line), the 
transition is always of the first order.} \la{xyc}
\end{figure}

The theory in \eq\nr{SMlagrangian} has a first 
order phase transition for small Higgs masses 
(small values of~$x$)~\cite{krreview}.
The transition gets weaker for larger Higgs masses,
and ends at $m_H\sim 80$ GeV~\cite{isthere}, 
see \fig\ref{xyc}.
Recently, the interest has been in studying
the endpoint region in some detail. Here, 
perturbation theory does not work at all and 
the dynamics is completely non-perturbative. 

The fact that there is an endpoint, was first 
reliably demonstrated in~\cite{isthere,karschnpr}. The endpoint
location was determined more precisely
in~\cite{gurtler}. A continuum 
extrapolation of the endpoint location was made in~\cite{endpoint},
employing improvement formulas derived in~\cite{moore_a}:
\be
x_c=0.0983(15), \quad
y_c=-0.0175(13).
\la{endloc}
\ee
In~\cite{endpoint}, it was also shown that the endpoint
belongs to the 3d Ising universality class.

\begin{table*}[hbt]
\setlength{\tabcolsep}{1.5pc}
\newlength{\digitwidth} \settowidth{\digitwidth}{\rm 0}
\catcode`?=\active \def?{\kern\digitwidth}
\caption{The endpoint location in different theories. 
No dimensional reduction error estimates
have been added to the 3d results here.
The values cited for $\sin^2\theta_W=0.23$ assume
that the finite volume and lattice spacing
effects do not depend on $\sin^2\theta_W$, see the text.
The comparison of critical temperatures between 4d and 
3d is sensitive to the relation of $g^2_\rmi{$\msbar$}$
and $g^2_R$, which is not known, while 
$m_{H,c}$ is much less sensitive. }

\vspace*{0.1cm}

\label{tab:endpoint}
\begin{tabular*}{\textwidth}{@{}l@{\extracolsep{\fill}}rrrrr}
\hline
 Method & Theory & \multicolumn{2}{c}{Couplings} &
                   \multicolumn{2}{c}{Endpoint location} \\
        &        & $\sin^2\theta_W$ &  $g^2_\rmi{$\msbar$}(m_W)$ & 
$m_{H,c}/$GeV &  $T_c(m_{H,c})/$GeV \\
\hline
 3d & Standard Model & $0$   &  $0.426$
 & 72.2(7) & 110.1(8) \\
    & Standard Model & $0.23$ & $0.426$
 & 72.3(7) & 109.2(8) \\ 
    & SU(2)+Higgs    & $0$    & $0.57(2)$ 
 & 65.9(7) & 133(4) \\
 4d~\cite{cfh3} & SU(2)+Higgs    & $0$    & $g^2_R\approx 0.57(2)$ 
 & 66.5(14)& 128(6) \\
\hline
\end{tabular*}
\end{table*}

The values in \eq\nr{endloc} can be converted to the endpoint
locations in different 4d physical theories, using the relations
derived in~\cite{generic}. Some values are given in Table~1.
The errors here represent the errors in \eq\nr{endloc}: no
additional errors 
have been added from dimensional reduction.

With 4d simulations, the endpoint location in the SU(2)+Higgs model
has been studied at a fixed (symmetric) lattice spacing
in~\cite{aoki1,aoki2}, and with an asymmetric lattice 
spacing in~\cite{cfh2,cfh3}. A continuum extrapolation has 
been carried out in~\cite{cfh3}, and that result is shown in Table~1.
It should be noted that the exact $\msbar$ gauge coupling to 
which the 4d simulations correspond, is not known. This affects
strongly the critical temperature ($T_c\propto m_H/g$), 
while the endpoint location itself is not that sensitive.

We can now compare the
3d and 4d results for SU(2)+Higgs.
Clearly, they are completely compatible.

Finally, consider the effect of $\sin^2\theta_W$. In general, 
the hypercharge U(1) group makes the transition slightly stronger,
though not by very much~\cite{su2u1}. Thus one might also expect
that the endpoint location changes to somewhat larger $x$ than in 
\eq\nr{endloc}. The infinite volume and 
continuum extrapolation of the endpoint
location has not been determined with $\sin^2\theta_W=0.23$, 
but finite volumes have been studied in~\cite{bext}.
On a lattice with $4/(g_3^2 a)=8$ and volume $=32^3$, we get
\ba
& & \hspace*{-0.7cm} x_c^0 = 0.1043(22),\quad y_c^0 = -0.02860(99), \nn
& & \hspace*{-0.7cm} x_c^1 = 0.1045(14),\quad y_c^1 = -0.02125(76),
\ea 
where ($^0$) refers to $\sin^2\theta_W=0$ 
and ($^1$) to $\sin^2\theta_W=0.23$.
Hence $x_c$ does not appear to depend 
significantly on  $\sin^2\theta_W$,
while $y_c$ changes a bit. Assuming that the same pattern remains there
at the infinite volume and continuum limits, the endpoint
location in physical units is given in Table~1 
also for $\sin^2\theta_W=0.23$. 

Recent topics of interest, other than the endpoint location, 
include the excitation spectrum 
around the endpoint~\cite{iss}, and
the behaviour of some topology related observables~\cite{cgis}. 

All in all, we can summarize the main lattice results 
for the Standard Model as follows:

\noindent
{\bf Perturbation theory:}
The non-perturbative transition is weaker
than in perturbation theory and
ends at $m_H=72(2)$ GeV.

\noindent
{\bf Cosmology:}
Although in principle all the ingredients 
for baryogenesis are there already
in the Standard Model, in practice one needs something more, 
to have a first order transition. 

\noindent
{\bf Dimensional reduction:}
In the first order regime, the results 
from 4d and 3d have been observed to 
be completely compatible~\cite{4d3d}. Now
a similar agreement has been demonstrated in
the non-perturbative endpoint regime~\cite{cfh3}. 
Thus we can be confident that
the accuracy estimates of dimensional reduction 
are reliable. This is important since theories
with chiral fermions can presently
only be studied with the 3d approach.

\subsection{MSSM}

In contrast to the Standard Model, baryogenesis could 
in principle work in the MSSM. This is because 
(1) there can be more CP-violation in the MSSM, 
due to new complex phases in the scalar sector
(see Sec.~\ref{sec:CP}), and because (2) 
there can be a stronger transition in the MSSM, due to a larger number of
light bosonic degrees of freedom, {\em viz.} the squarks.
Thus MSSM is a natural candidate for electroweak baryogenesis. 

The parameter regime relevant for MSSM baryogenesis has  
been studied extensively in perturbation theory.
The relevant case has been found to be 
that the right-handed stops are light, 
$m_{\tilde t_R} < m_{\rm top}$, and the Higgs mass is anything
allowed by experiment and by the MSSM, 
$m_H=$ 90...110 GeV~[24--27].

As was the case
for the Standard Model, perturbative estimates are nevertheless
not necessarily reliable. Thus, the transition has been studied
with 3d simulations~\cite{mssmsim}.

The light degrees of freedom appearing in the 3d effective theory
are now the spatial components of the SU(2) and SU(3)
gauge fields, the right-handed
stop $U$, and the combination $H$ of the two Higgses 
$H_1,\tilde H_2$ which is light
at the phase transition point. The corresponding action is
\ba
& & \!\!\!\!\!\!\!\!\!\! {\cal L}_{\rm 3d} =
\fr14 F^a_{ij}F^a_{ij}+\fr14 G^A_{ij}G^A_{ij}+ 
\gamma_3 H^\dagger H U^\dagger U \la{HHUUaction} \\
& & + (D_i^w H)^\dagger(D_i^w H)+m_{H3}^2 H^\dagger H+
\lambda_{H3} (H^\dagger H)^2 \nn 
& & + (D_i^s U)^\dagger(D_i^s U)+m_{U3}^2U^\dagger U+
\lambda_{U3} (U^\dagger U)^2. \nonumber
\ea
Here $D_i^w,D_i^s$ are the SU(2) and SU(3) covariant 
derivatives, and $F_{ij}^a, G_{ij}^A$ are the 
corresponding field strengths.
The parameters of this action can be determined
in the standard way~\cite{mssmsim}.

To simulate the theory in \eq\nr{HHUUaction}, 
is straightforward but technically demanding. 
For details, we refer to~\cite{mssmsim}.

The basic result of the simulations is shown in 
\fig\ref{v2v3}. There we show the Higgs field expectation
value after the transition, for $m_H=95$ GeV, as 
a function of a parameter $\tilde m_U$ which 
determines the zero temperature right-handed stop mass through 
$m_{\tilde t_R} \approx ({m_{\rm top}^2-\tilde m_U^2})^{1/2}
\approx 160...150$ GeV.
The lattice results are compared with 
2-loop perturbation theory (the solid line).

\begin{figure}[tb]

\vspace*{0.2cm}

\epsfxsize=7.3cm
\centerline{\epsffile{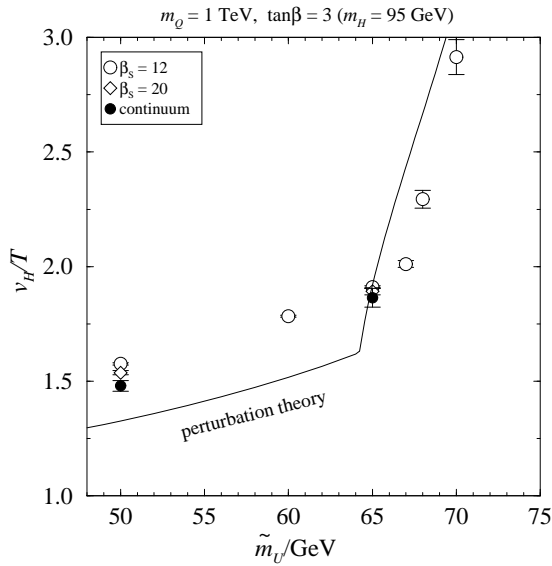}}
 
\vspace*{-1cm}

\caption[a]{The Higgs field expectation value
in the MSSM~\cite{mssmsim}, compared with 2-loop perturbation
theory. For $\tilde m_U\gsim 64...67$ GeV, 
there is a two-stage transition,
and the values of $v_H/T$ shown represent
the situation after the latter stage.}
\la{v2v3}
\end{figure}

Based on \fig\ref{v2v3}, we can summarize the lattice
results for the MSSM electroweak phase transition as follows:

\noindent
{\bf Perturbation theory:}
In contrast to the Standard Model, the
electroweak phase transition in the MSSM 
can be stronger than in 2-loop perturbation theory,
even for large $m_H$
(although the difference is not very large).
Thus perturbation theory gives a conservative estimate. 

\noindent
{\bf Cosmology:}
If the non-perturbative strengthening effect remains
there also for larger Higgs masses than shown in \fig\ref{v2v3}, 
then Higgs masses up to 105...110~GeV are allowed
for baryogenesis, provided 
that $m_{\tilde t_R} < m_{\rm top}$~\cite{cqw2,cm}.

\section{FUTURE PROSPECTS}
\la{sec:CP}

As we have seen, lattice simulations have been quite successful
in solving the problem of the electroweak phase transition:
the case of the Standard Model is now completely understood, 
and similar techniques have been applied
also in the MSSM. One has also been able to demonstrate
with 4d simulations that the dimensionally reduced 3d theory
works for non-perturbative quantities as well as for
perturbative ones: the non-perturbative infrared dynamics 
is really three-dimensional, as it has to be (see also~\cite{jl}). 

What is it then that still remains to be done?
We list here a few open questions,
related in particular to the MSSM:

1. What happens for other 
parameter values in the MSSM
(a larger $m_H$, non-vanishing
squark mixing, etc), when the transition gets weaker?

2. Could 4d simulations be used to estimate non-perturbatively
the accuracy of dimensional reduction 
in a theory similar to the MSSM? Could 
4d simulations be used in some regions of the parameter space~\cite{gk}
where the theory is not weakly coupled at zero temperature, 
and thus dimensional reduction does not work?

3. Finally, can one say something about 
CP-violation with simulations? The existence of CP-violation
is one of the ingredients for electroweak baryogenesis:
otherwise, one produces
the same amounts of baryons and anti-baryons, and no net
asymmetry arises. Let us discuss the last question in some more detail.

To get enough CP-violation for producing the observed
baryon asymmetry, turns out to be a non-trivial requirement~\cite{rs}. 
It is hence 
an interesting prospect that there are new sources
of CP-violation in the MSSM, related
to the trilinear couplings of Higgses and the
squarks.

An intriguing observation is now that since 
one has two Higgses in the MSSM, the effect of the new
CP-violating parameters can also propagate to 
a CP-violating
phase between the Higgses. Thus, the idea arises that maybe the
there is a non-trivial profile of the CP-violating phase
at the phase boundary between the broken and symmetric
phases~[31--33].  
The phase boundary is the region relevant for 
baryogenesis, and this scenario might allow for 
sufficient CP-violation, without violating the constraints 
that have to be satisfied in the broken phase. 

It appears that whether such a phenomenon
takes place, can again be studied non-pertur\-ba\-tively
with a simple 3d theory.
The theory in \eq\nr{HHUUaction} is not enough, though:
in that case, we were interested in the strength of the 
transition, and it was sufficient to study a particular
linear combination of $H_1, \tilde H_2$.
Now we are interested in a question for which
both of the Higgses $H_1,\tilde H_2$ should be kept in the 
effective theory. The effective theory will then 
be more complicated than in \eq\nr{HHUUaction},
but can be derived with precisely the same methods.
In such a theory, there are new CP-violating
operators, e.g.  $\im H_1^\dagger\tilde H_2$, which 
could have a non-trivial profile at the phase 
boundary between the symmetric and broken phases.
This problem could, in principle, 
be solved with lattice simulations.


\end{document}